\documentclass[prd,twocolumn,amsmath,amssymb,floatfix,superscriptaddress]{revtex4-1}
\usepackage{graphicx}
\usepackage{amssymb}
\usepackage{amsmath}
\usepackage{float}
\usepackage{placeins}
\usepackage{color}
\usepackage{ wasysym }
\usepackage{hyperref}
\usepackage{tabu}
\usepackage{bm}
\usepackage{float}
\usepackage{xcolor}
\def\barray{\begin{array}}
\def\earray{\end{array}}
\def\be{\begin{equation}}
\def\ee{\end{equation}}
\def\ben{\begin{equation} \nonumber}
\def\een{\end{equation}}
\def\ban{\begin{eqnarray*}}
\def\ean{\end{eqnarray*}}
\def\ba{\begin{eqnarray}}
\def\ea{\end{eqnarray}}

\def\({\left(}
\def\){\right)}

\graphicspath{{./fig/}}

\begin{document}

\title{Does Early Dark Energy Absorb the DESI Late-Time Dynamics Signal? A Combined Analysis}
\author{Mohammad Malekjani}
\email{malekjani@iasbs.ac.ir}
    \affiliation{Department of Physics, Institute for Advanced Studies in Basic Sciences (IASBS), P.O. Box 45137-66731, Zanjan, Iran}
\author{Saeed Pourojaghi}
%\email{s.pouri90@gmail.com}
\affiliation{School of Physics, Institute for Research in Fundamental Sciences (IPM), P.O.Box 19395-5531, Tehran, Iran}

\begin{abstract}
While the recent Baryon Acoustic Oscillation (BAO) measurements from the Dark Energy Spectroscopic Instrument (DESI) collaboration are largely consistent with a flat $\Lambda$CDM cosmology, the preferred parameters are in mild tension with those determined from the cosmic microwave background (CMB). A late-time dynamical dark energy (DDE) solution has been proposed by the DESI collaboration to address this tension. In this work, we investigate whether the statistical preference for DDE is a genuine late-time phenomenon or an artifact of unresolved early-universe physics. To do so, we simultaneously allow for both early- and late-time modifications to the expansion history by combining the Early Dark Energy (EDE) framework with the Chevallier-Polarski-Linder (CPL) parametrization. Excluding the DESI BAO measurements, our joint analysis of the CMB+Pantheon+ datasets demonstrates that within an EDE-extended framework, the CPL parameters remain statistically consistent with the standard $\Lambda$CDM model. This supports the hypothesis that a DDE signal at low redshifts can be effectively accounted for by an EDE component within the $\Lambda$CDM background. However, upon the inclusion of the DESI BAO measurements in the joint analysis, a statistically significant deviation from a cosmological constant emerges. Within this combined framework, the best-fit CPL parameters robustly indicate a departure from the standard $\Lambda$CDM model, favoring a phantom-to-quintessence transition in the DE equation of state. This demonstrates that the DESI preference for the late-time DDE is a robust signature that cannot be absorbed by modifying the physics of the early Universe.
\end{abstract}
\maketitle

%%%%%%%%%%%%%%%%%%%%%%%%%%%%%%%%%%%%%%%%%%%%%%%%%%%%%%%%%%%%%%%%%%%%%%%%%%%%%%%%%%%%%%%%%%%%%
\section{Introduction} \label{sec:intro}
The discovery of the late-time accelerated expansion of the Universe, primarily inferred from Type Ia Supernovae (SNeIa) observations over a quarter of a century ago \cite{Riess:1998cb,Perlmutter:1998np}, fundamentally changed our understanding of cosmic dynamics. To account for this paradigm-shifting phenomenon, Einstein's cosmological constant, $\Lambda$, was introduced. Together with Cold Dark Matter (CDM) in a spatially flat geometry, $\Lambda$ establishes the well-known standard flat $\Lambda$CDM cosmological model, providing a highly successful and remarkably simple framework that robustly explains the accelerated expansion phase discovered by the SN Ia data \cite{Riess:1998cb,Perlmutter:1998np}.

This initial evidence for late-time cosmic acceleration has since been robustly corroborated by a wealth of independent cosmological probes. Most notably, the latest extensive SN Ia data, such as the Pantheon+, Union3 and the Dark Energy Survey (DES)-Dovekie samples \cite{Scolnic:2021amr, Rubin:2023ovl, DES:2025sig}, high-precision measurements of the Cosmic Microwave Background (CMB) anisotropies \cite{WMAP2013,Planck:2013pxb,Planck:2018vyg,Planck:2019nip} and the Baryon Acoustic Oscillations (BAO) imprinted in the Large Scale Structure (LSS) provide strong complementary support. Traced by extensive cosmological surveys such as Sloan Digital Sky Survey
(SDSS), Baryon Oscillation Spectroscopic Survey (BOSS), extended BOSS (eBOSS), and most recently the DESI BAO scale act as a standard cosmic ruler that independently measures the expansion history. These observations consistently point to a phase of cosmic acceleration and reinforce the flat $\Lambda$CDM paradigm \cite{Eisenstein:2005su,Alam2017, DES:2021wwk,DESI:2024mwx,DESI:2025zgx}. In addition to the aforementioned evidence,  cosmological data from galaxy clusters and weak lensing surveys further corroborate this picture, affirming that the accelerated expansion scenario is a cornerstone of our current cosmological framework \cite{Vikhlinin2009, Heymans2013,DES:2021wwk,KiDS:2020suj,eROSITA:2024qzn}.

Despite these remarkable empirical successes, the standard $\Lambda$CDM cosmological model is increasingly challenged on both theoretical and observational fronts. Theoretically, the physical nature of the cosmological constant remains elusive, burdened by the well-known fine-tuning problem as well as the cosmic coincidence problem \cite{Weinberg:1988cp,Padmanabhan:2002ji}.
Observationally, as measurement precision has improved, persistent statistical tensions have surfaced between early- and late-universe probes. The most critical of these is the $H_0$ tension, an unresolved and severe discrepancy between the present-day expansion rate inferred from the CMB under $\Lambda$CDM model ($H_0 = 67.4 \pm 0.5 \, \mathrm{km \, s^{-1} \, Mpc^{-1}}$ \cite{Planck:2018vyg}) and that measured by local distance-ladder calibrations (yielding values around $H_0 = 73.0 \pm 1.0 \, \mathrm{km \, s^{-1} \, Mpc^{-1}}$ \cite{Riess2019,Riess:2021jrx,Freedman2021}). This is accompanied by other mild discrepancies, such as the $\sigma_8$ tension, which highlights a mismatch in the amplitude of matter clustering predicted by the standard model versus local structural observations \cite{Pergola2023Tensions}. A comprehensive overview of these phenomenological tensions can be found in recent reviews \cite{Perivolaropoulos:2021jda}.

Over the past decades, a wide variety of DDE models have been proposed to address, or at least alleviate, the theoretical challenges associated with a rigid cosmological constant. Among the most prominent physically motivated alternatives are quintessence models, which typically invoke a canonical, slowly rolling scalar field to drive the late-time acceleration of the Universe \cite{Ratra:1987rm,Caldwell:1997ii,Tsujikawa:2013fta}. Unlike the cosmological constant, whose equation of state (EoS) is fixed at $w_{\Lambda} = -1$, these dynamical frameworks allow the EoS to evolve with cosmic time, thereby providing a natural avenue to circumvent the fine-tuning and cosmic coincidence problems. 
To test such DDE scenarios against observations in a largely model-independent way, phenomenological parameterizations of the EoS are widely employed. Among them, the most commonly used and extensively studied form is the Chevallier--Polarski--Linder (CPL) parametrization \cite{Chevallier:2000qy,Linder:2002et}, $w_{\rm DE}(a) = w_0 + w_a(1-a)$, where $a$ denotes the cosmic scale factor. The CPL parametrization offers a simple yet powerful two-parameter description, $(w_0, w_a)$, capable of capturing both the present-day value of the DE EoS and its possible time evolution.

While simple late-time modifications to the expansion history of the Universe such as DDE are sufficient to address the theoretical problems of $\Lambda$CDM cosmology, they are insufficient to fully resolve the $H_0$ tension. In this regard, alternative theoretical pathways are generally proposed in two broad categories: introducing interactions within the dark sector at late times \cite{Gomez-Valent:2026ept}, such as in interacting dark energy-dark matter (IDE) models \cite{Wang:2016lxa, DiValentino:2019ffd, Lucca:2020zjb,Li:2026xaz,vanderWesthuizen:2025mnw,vanderWesthuizen:2025rip,Pan:2025qwy}, or modifying the pre-recombination physics to reduce the sound horizon at the drag epoch ($r_d$). Among the early-universe solutions, Early Dark Energy (EDE) has emerged as one of the most successful and extensively studied frameworks \cite{Poulin:2018cxd, Kamionkowski:2022pkx}. The EDE model postulates the existence of a transient scalar field that briefly contributes a significant fraction to the total energy density of the Universe shortly before recombination, naturally decreasing $r_d$ and thereby yielding a higher inferred value of $H_0$ in agreement with local measurements \cite{Du:2026qtq}.

While individual cosmological probes of the background expansion---including the CMB and CMB lensing \cite{Planck:2018vyg,ACT:2023kun,SPT-3G:2023rvw}, SNeIa compilations (Pantheon+ \cite{Scolnic:2021amr}, Union3 \cite{Rubin:2023ovl}, DES-SN5YR \cite{DES:2024tys}, DES-Dovekie \cite{DES:2025sig}), and BAO measurements from DESI \cite{DESI:2024mwx, DESI:2025zgx}---are broadly consistent with a flat-$\Lambda$CDM cosmology, a mild but intriguing tension emerges in their cross-correlation. Specifically, high-precision BAO measurements exhibit slight deviations from the distance-redshift relations predicted by the Planck-calibrated $\Lambda$CDM baseline. In this context, the groundbreaking BAO measurements recently released by the DESI collaboration map the cosmic expansion history with unprecedented precision \cite{DESI:2024mwx, DESI:2025zgx}. The combination of DESI BAO data with early-universe CMB observations and recent SNeIa compilations reveals a compelling statistical preference for DDE over a rigid cosmological constant (for a review, see also \cite{Li:2026asg}). Most notably, within the CPL parameterization, this joint analysis favors an evolving EoS for DE characterized by $w_0 > -1$ and $w_a < 0$. Physically, this specific parameter space indicates a striking cosmological feature: a transition of EoS from an early phantom regime ($w_{\rm DE}(z) < -1$) at higher redshifts to a present-day quintessence phase ($w_0 > -1$). Interestingly, this characteristic transition is consistently recovered even when employing alternative two-parameter $w_0w_a$ parametrizations \cite{Malekjani:2024bgi,Giare:2024gpk}.
This evolving DE signal underscores the critical need to confront combined early- and late-time datasets to uncover potential dynamics in the dark sector. Furthermore, the phantom-to-quintessence transition inferred from the DESI observations carries possible implications for the ultimate fate of the Universe. The crossing of the phantom divide towards $w_{\rm DE} > -1$ might serve as a compelling hint for the eventual end of the dark energy-dominated era. If this dynamical trend continues, the cosmic acceleration could weaken and eventually cease, drastically altering the standard de Sitter future predicted by $\Lambda$CDM \cite{Li:2026hwq}.

Recently, EDE has been proposed as an early-time modification to the expansion history, serving as a compelling alternative to late-time DDE scenarios\cite{Chaussidon:2025npr}. Recent analyses demonstrate that an EDE cosmology with a flat-$\Lambda$CDM late-time background provides a significantly better overall fit to current cosmological data combinations than late-time DDE modifications based on the CPL parameterization \cite{Chaussidon:2025npr}. This preference is largely driven by EDE's essential role in resolving the $H_0$ tension when local distance-ladder calibrators are included.

In particular, when SNeIa calibrated by Cepheids are included in the Pantheon+ catalog (often referred to as Pantheon+SH0ES), the combination of CMB, DESI BAO, and Pantheon+SH0ES strongly prefers EDE over late-time DDE. The reason is clear: EDE successfully mitigates the $H_0$ tension, thus accommodating the SH0ES data much better than late-time DDE scenarios. 
Given that EDE scenarios already offer a superior fit to the combined high-redshift CMB and low-redshift (BAO and Pantheon+SH0ES) data, it becomes crucial to determine whether any modification to the late-time cosmology remains necessary. It should be noted that the analysis in \cite{Chaussidon:2025npr} was limited to a direct comparison between a $\Lambda$CDM supplemented by EDE model and a standalone late-time CPL modification. In a more general framework, we aim to consider early- and late-time modifications to the Hubble flow (EDE and CPL) and compare the results with EDE and CPL individually. By performing this comprehensive comparison, we are able to answer whether the late-time DDE signal proposed by recent DESI results persists within an EDE cosmology. For this analysis, the datasets we use are Planck CMB \cite{Planck:2019nip}, DESI BAO DR2 \cite{DESI:2025zgx}, and the SNeIa Pantheon+SHOES catalog \cite{Scolnic:2021amr}.
The structure of the paper is organized as follows: In Section (\ref{sect:models}), we present the theoretical frameworks of EDE and late-time DE modifications to the Hubble flow. Section (\ref{sect:cosm_models}) introduces our cosmological models and briefly explains the methodology and observational data we have used.  In Section (\ref{sect:num_results}), we present our numerical results for different DE parameterizations considered in our analysis. Finally, in Section (\ref{sect:conclusion}), we summarize our findings and conclude the study.

%%%%%%%%%%%%%%%%%%%%%%%%%%%%%%%%%%%%%%%%%%%%%%%%%%%%%%%%%%%%%%%%%%%%%%%%%%%%%%%%%%%%%%%%%%%
\section{Theoretical Framework: Early and Late Dark Energy Dynamics} \label{sect:models}
In this section, we introduce the theoretical frameworks adopted in this work to explore modifications to the expansion history of the Universe. We begin with the EDE scenario, which introduces a localized contribution to the Hubble rate in the pre-recombination era. We then discuss the CPL parametrization, which allows for DDE at late times.

\subsection{EDE Dynamics}\label{sect:subsction:2-1}
The core idea of EDE is to introduce a new cosmological component that is dynamically relevant in the pre-recombination era. Typically, this component is modeled as a scalar field, denoted by $\phi$, whose background dynamics are simply governed by the homogeneous Klein-Gordon equation. At very early times, the field is held nearly constant by Hubble friction, such that its energy density is approximately constant. As the Universe expands and the Hubble parameter $H(z)$ drops, the friction term becomes subdominant; the field is then released from its frozen state and eventually becomes dynamical. Unlike during the frozen phase, the energy density of the field now rapidly dilutes, redshifting away faster than matter or radiation. This brief, localized contribution to the Hubble flow provides a suitable solution to reduce the sound horizon and can thus resolve the Hubble tension \cite{Poulin:2018cxd}.
To capture the essential physics of an EDE scalar field without specifying the exact form of the potential, a phenomenological fluid model is often adopted. In this work, we adopt the \textit{Acoustic Early Dark Energy (ADE)}~\cite{Lin:2019fqo} to parameterize the background behavior of the EDE fluid through its EoS, $w_{\rm EDE}(a) \equiv p_{\rm EDE}/\rho_{\rm EDE}$, as:
\begin{equation}
w_{\rm EDE}(a) = \frac{1 + w_f}{1 + (a_c / a)^{3(1 + w_f)}} - 1, \label{eq:pheno_w}
\end{equation}

where $a$ is the scale factor. This functional form describes a fluid that behaves like a cosmological constant ($w_{\rm EDE} \simeq -1$) at very early times ($a \ll a_c$), causing its energy density to be nearly constant. The transition to a dynamical regime occurs at around the critical scale factor $a_c$, after which the fluid dilutes like a species with an EoS $w_f$, such that $\rho_{\rm EDE} \propto a^{-3(1+w_f)}$ for $a \gg a_c$. In the context of a canonical scalar field oscillating in a potential $V \propto \phi^{2n}$ around its minimum, the final EoS is given by $w_f = (n-1)/(n+1)$, which lies in the range $0 \leq w_f \leq 1$~\cite{Poulin:2018dzq,Turner:1983he}. However, for non-canonical scalar fields, $w_f$ can exceed unity, leading to a faster dilution of the EDE component. This is significant because larger values of $w_f$ allow for a larger fractional contribution $f_{\rm EDE}$ at the critical scale factor $a_c$, as the EDE can provide a more substantial boost to the Hubble rate while still diluting away sufficiently quickly after the transition.\\
Given the EoS in Eq.~\eqref{eq:pheno_w}, the background energy density of the EDE fluid evolves as:
\begin{equation}
\rho_{\rm EDE}(a) = \rho_{\rm EDE}(a_c) \exp \left[ -3 \int_{a_c}^{a} \frac{1 + w_{\rm EDE}(a')}{a'} da' \right], \label{eq:rho_ede}
\end{equation}

Consequently, the Hubble parameter in the EDE-$\Lambda$ cosmology (indicating EDE as the primary modification, with $\Lambda$ at late times) can be written as:
\begin{equation}
H^2(a) = H_0^2 \left[ \Omega_{\rm r} a^{-4} + \Omega_{\rm m} a^{-3} + \Omega_{\rm EDE}(a) + \Omega_{\rm \Lambda} \right], \label{eq:Huuble}
\end{equation}

where $\Omega_{\rm EDE}(a) \equiv \rho_{\rm EDE}(a)/\rho_{\rm crit}(a)$ captures the localized contribution of the EDE component around the critical scale factor and can be written as:
\begin{equation}
\Omega_{\rm EDE}(a) = \frac{f_{\rm EDE}}{1-f_{\rm EDE}} [\Omega_{\rm r} a_c^{-4} + \Omega_{\rm m} a_c^{-3}] \frac{2a_c^n}{a^n+a_c^n}.
\label{eq:Omega_ede}
\end{equation}

In the context of the phenomenological EDE parameterization, we are allowed to vary the key parameters---the maximum fractional energy density contribution $f_{\rm EDE} \equiv \rho_{\rm EDE}/\rho_{\rm tot}$ at $a_c$, the critical redshift $z_c=\frac{1}{a_c}-1$ itself, and the post-transition EoS $w_f$---to explore their impact on cosmological observables. This minimal phenomenological model serves as a powerful tool to understand the phenomenological requirements for resolving the Hubble tension, as it captures the essential background dynamics common to many specific scalar-field models~\cite{Lin:2019fqo}.

\subsection{Late-Time Dynamics via CPL Parameterization}\label{sect:subsction:2-2}
While EDE modifies the expansion history in the pre-recombination era, a conceptually different approach is to consider modifications to the expansion history at late times ($z \lesssim 2$). Instead of a cosmological constant $\Lambda$, one can allow the EoS of DE to evolve with redshift. This can alter the late-time expansion rate $H(z)$ and the distance-redshift relation.
The most widely used framework for exploring this possibility is the CPL parameterization ~\cite{Chevallier:2000qy,Linder:2002et} given by:
\begin{equation}
w_{\rm DE}(z) = w_0 + w_a (1-a), \label{eq:cpl}
\end{equation}

where $w_0$ represents the present-day value of the EoS, and $w_a$ governs its time evolution. This functional form provides a simple two-parameter description of DDE that remains finite at high redshift ($a \to 0 $, $w \to w_0 + w_a$)~\cite{Chevallier:2000qy,Linder:2002et}.
In a flat Universe, the Hubble parameter in the framework of DE cosmology is simply given by:
\begin{equation}
H^2(z) = H_0^2 \left[ \Omega_m (1+z)^3 + (1-\Omega_m) \, f_{\rm DE}(z) \right],
\end{equation}

where the normalized DE density $f_{\rm DE}(z)$ is determined by the EoS through
\begin{equation}
f_{\rm DE}(z) = \exp \left( 3 \int_0^z \frac{1 + w_{\rm DE}(z')}{1+z'} dz' \right).
\end{equation}

Substituting the CPL parametrization yields the analytical expression~\cite{Linder:2002et}:
\begin{equation}
f_{\rm DE}(z) = (1+z)^{3(1+w_0+w_a)} \exp\left(-\frac{3 w_a z}{1+z}\right).
\end{equation}

 Depending on the values of $w_0$ and $w_a$, the CPL model can produce different behaviors---ranging from quintessence-like ($w_{\rm DE}> -1$) to phantom-like ($w_{\rm DE}< -1$) DE. The recent DESI BAO data \cite{DESI:2024mwx,DES:2025sig} togather Planck CMB \cite{Planck:2019nip} and SNeIa compilations \cite{Scolnic:2021amr,Rubin:2023ovl,DES:2025sig} favor $w_0 > -1$ and $w_a < 0$, which corresponds to an EoS of DE that evolves from a phantom-like behavior at early times to a quintessence-like behavior at late times~\cite{DESI:2024mwx,DESI:2025zgx}. 
 This signal could be an indication of new physics beyond the standard cosmological model. However, the statistical significance and robustness of this preference remain subjects of active debate in the literature~\cite{Nesseris:2025lke,Malekjani:2025alf}. In this regard, future surveys and improved analysis techniques will be crucial in determining whether this signal persists or is merely a statistical fluctuation. Notice that prior to the DESI results, a mild indication of evolving dark energy was reported in \cite{Brout:2022vxf} from the combination of Pantheon+, Planck CMB, and BAO data within a flat $w_0w_a$CDM framework. However, the departure from the $\Lambda$CDM point $(w_0,w_a)=(-1,0)$ was only at the $1\sigma$ level, while $\Lambda$CDM remained well within the $2\sigma$ confidence region. Consequently, the BAO data before DESI measurements did not provide statistically significant evidence for evolving DE.

%%%%%%%%%%%%%%%%%%%%%%%%%%%%%%%%%%%%%%%%%%%%%%%%%%%%%%%%%%%%%%%%%%%%%%%%%%%%%%%%%%%%%%%%%%%%%
\section{Cosmological models and methodology}\label{sect:cosm_models}
\subsection{Cosmological Models}
In this work, we consider four cosmological models to investigate the impact of modifications to the expansion history on cosmological observables. These models are summarized in Table~\ref{tab:models} and described below.
\begin{table*}
\centering
\begin{tabular}{|c|c|c|}
\hline
\textbf{Model} & \textbf{Abbreviation} & \textbf{Description} \\
\hline
$\Lambda$CDM  & $\Lambda$CDM & Standard cosmological model with cosmological constant $\Lambda$ \\
CPL           & CPL          & DDE via CPL parameterization at late times \\
$\Lambda$CDM augmented by EDE& EDE-$\Lambda$CDM & EDE with cosmological constant $\Lambda$ at late times \\
CPL augmented by EDE & EDE-CPL & EDE with DDE via CPL parametrization at late times \\
\hline
\end{tabular}
\caption{Summary of the cosmological models analyzed in this work.}
\label{tab:models}
\end{table*}

The free parameters for the base flat-$\Lambda$CDM cosmology are the baryon density $\omega_b = \Omega_b h^2$, the total matter density $\omega_{\rm m} = \Omega_{\rm m}h^2$, the Hubble constant $h$ (or $H_0 = 100 h$) and the absolute magnitude of SNeIa $M$. For the second model, CPL has two additional free parameters, $w_0$ and $w_a$, compared to standard flat-$\Lambda$CDM cosmology.
The third model, EDE-$\Lambda$CDM, augments $\Lambda$CDM with an EDE component as described in Sec.~\ref{sect:subsction:2-1}. This adds three free parameters: the maximum fractional contribution of EDE at the critical redshift $f_{\rm EDE}$, the critical redshift $z_c$ at which the EDE becomes dynamical, and the post-transition EoS $w_f$ (or equivalently, the exponent $n$). For this model, the late-time DE remains a cosmological constant.
The fourth model, EDE-CPL, combines the EDE component at early times with DDE at late times through the CPL parametrization. This model contains the full set of parameters from both EDE (three parameters) and CPL (two parameters), in addition to the standard $\Lambda$CDM parameters. The EDE-CPL model allows us to study whether a combined early- and late-time modification of the expansion history can provide a better fit to the data than either modification alone.

\subsection{Data and Methodology}
In this section, we briefly describe the observational data and statistical methods employed in our analysis. We consider various combinations of Pantheon+ SNeIa, DESI BAO, and Planck CMB data to constrain the parameters of the cosmological models under consideration. The late-time expansion history is probed using the Pantheon+ SNeIa sample \cite{Scolnic:2021amr}, while distance measurements at intermediate redshifts are provided by the DESI DR2 BAO data \cite{DESI:2025zgx}. We also incorporate compressed CMB data through distance-prior constraints, which capture the principal geometrical information relevant to background-level cosmological analyses \cite{Zhai:2018vmm}.
 
The Pantheon+ compilation provides a homogeneous sample of standardized SNeIa spanning the redshift range $0.001 < z < 2.26$ \cite{Scolnic:2021amr, Brout:2022vxf}. For a spatially flat geometry, the luminosity distance and the corresponding theoretical distance modulus are given by
\begin{equation}
D_L(z;\boldsymbol{\theta})
=
c(1+z)\int_0^z \frac{dz'}{H(z';\boldsymbol{\theta})},
\end{equation}
and
\begin{equation}
\mu_{\rm th}(z;\boldsymbol{\theta})
=
5\log_{10}
\left[
\frac{D_L(z;\boldsymbol{\theta})}{\mathrm{Mpc}}
\right]
+25,
\end{equation}

where $\boldsymbol{\theta}$ denotes the set of cosmological parameters.

Among the Pantheon+ SNeIa, 77 are calibrated using Cepheid-host galaxies. We refer to these as the SH0ES SNeIa. In our analysis, we separate this subsample from the remaining Hubble-flow SNeIa and construct an independent likelihood for each subsample. For the Hubble-flow sample, the corresponding chi-square function is
\begin{equation}
\chi^2_{\rm SN}
=
\left(\boldsymbol{\Delta}^{\rm SN}\right)^{T}
\mathbf{C}_{\rm SN}^{-1}
\boldsymbol{\Delta}^{\rm SN},
\end{equation}
where
\begin{equation}
\Delta_i^{\rm SN}
=
m_i - M - \mu_{\rm th}(z_i;\boldsymbol{\theta}).
\end{equation}

Here, $m_i$ denotes the observed apparent magnitude of the $i$th Hubble-flow SNeIa, and $M$ is the standardized absolute magnitude.

For the SH0ES subsample, we use
\begin{equation}
\chi^2_{\rm SH0ES}
=
\left(\boldsymbol{\Delta}^{\rm SH0ES}\right)^{T}
\mathbf{C}_{\rm SH0ES}^{-1}
\boldsymbol{\Delta}^{\rm SH0ES},
\end{equation}
where
\begin{equation}
\Delta_j^{\rm SH0ES}
=
m_j - M - \mu_j^{\rm SH0ES}.
\end{equation}

In this expression, $\mu_j^{\rm SH0ES}$ is the independently measured distance modulus of the host galaxy of the $j$th SH0ES SN Ia, while $m_j$ is its observed apparent magnitude. The covariance matrices $\mathbf{C}_{\rm SN}$ and $\mathbf{C}_{\rm SH0ES}$ are obtained by restricting the full Pantheon+ covariance matrix to the corresponding sub-samples.

The DESI DR2 BAO data \cite{DESI:2025zgx} used in this analysis consist of both anisotropic and isotropic measurements. Anisotropic BAO measurements retain separate information about distances transverse and parallel to the line of sight. They therefore constrain the transverse comoving distance, $D_M(z)$, and the radial distance scale, $D_H(z)=c/H(z)$, independently. The corresponding data vector contains twelve entries, arranged in pairs at six effective redshifts,
\begin{equation}
    \mathbf{D}_{\rm aniso}
    =
    \left(
    \frac{D_M(z_1)}{r_d},
    \frac{D_H(z_1)}{r_d},
    \ldots,
    \frac{D_M(z_6)}{r_d},
    \frac{D_H(z_6)}{r_d}
    \right),
\end{equation}

where
\begin{equation}
    z_i =
    0.51,\ 0.706,\ 0.934,\ 1.321,\ 1.484,\ 2.33.
\end{equation}

Here, $D_M(z)$ is the transverse comoving distance, $D_H(z)=c/H(z)$ is the Hubble distance, and $r_d$ is the comoving sound horizon at the baryon drag epoch. For a spatially flat cosmological model, the theoretical quantities used in the analysis are
\begin{equation}
    \begin{aligned}
        \frac{D_M(z)}{r_d}
        &=
        \frac{c}{H_0 r_d}
        \int_0^z \frac{dz'}{E(z')}, \\
        \frac{D_H(z)}{r_d}
        &=
        \frac{c}{H_0 r_d E(z)} .
    \end{aligned}
\end{equation}

where $E(z)=H(z)/H_0$.

In our analysis, $r_d$ is not fixed to an external Planck value and is not treated as an independent free parameter. Instead, at each point in the cosmological parameter space, it is computed self-consistently from the sound-horizon integral at the drag epoch,
\begin{equation}
    r_d
    =
    \int_{z_d}^{\infty}
    \frac{c_s(z)}{H(z)}\,dz,
\end{equation}

where $z_d$ is the baryon drag redshift and $c_s(z)$ is the sound speed in the tightly coupled photon--baryon fluid. Therefore, the BAO observables are calibrated internally within each cosmological model rather than by imposing an external prior on $r_d$.

The covariance matrix of the anisotropic data is constructed from the quoted uncertainties and the corresponding correlation matrix. So we can write the chi-square function as
\begin{equation}
    \chi^2_{\rm aniso}
    =
    \Delta\mathbf{D}_{\rm aniso}^{T}
    \mathbf{C}_{\rm aniso}^{-1}
    \Delta\mathbf{D}_{\rm aniso},
\end{equation}

where
\begin{equation}
    \Delta\mathbf{D}_{\rm aniso}
    =
    \mathbf{D}_{\rm aniso}^{\rm obs}
    -
    \mathbf{D}_{\rm aniso}^{\rm th}.
\end{equation}

In contrast, isotropic BAO measurements compress the transverse and radial distance information into a single volume-averaged distance. In addition to the anisotropic measurements, we include one isotropic BAO measurement at $z=0.295$, expressed as
\begin{equation}
    \frac{D_V(z)}{r_d}
    =
    \left[
    \left(\frac{D_M(z)}{r_d}\right)^2
    \frac{zD_H(z)}{r_d}
    \right]^{1/3}.
\end{equation}

Assuming that this measurement is independent of the anisotropic BAO data, its chi-square is
\begin{equation}
    \chi^2_{\rm iso}
    =
    \left[
    \frac{
    (D_V/r_d)_{\rm obs}
    -
    (D_V/r_d)_{\rm th}
    }{
    \sigma_{D_V/r_d}
    }
    \right]^2.
\end{equation}

The total DESI BAO contribution is consequently given by
\begin{equation}
    \chi^2_{\rm BAO}
    =
    \chi^2_{\rm aniso}
    +
    \chi^2_{\rm iso}.
\end{equation}

Finally, we include compressed CMB information through a set of distance-prior constraints \cite{Zhai:2018vmm}. The data vector used in this analysis is
\begin{equation}
    \mathbf{v}_{\rm CMB} = \left(R,\ell_A,\Omega_b h^2,n_s\right),
\end{equation}

with the observed values
\begin{equation}
    \mathbf{v}_{\rm CMB}^{\rm obs}
    =
    \left(
    1.74963,\ 301.80845,\ 0.02237,\ 0.96484
    \right).
\end{equation}

Here $R$ is the CMB shift parameter, $\ell_A$ is the acoustic angular scale, $\Omega_b h^2$ is the physical baryon density, and $n_s$ is the scalar spectral index. The theoretical quantities are computed as
\begin{equation}
    R = \sqrt{\Omega_m}\, \frac{H_0 D_C(z_\ast)}{c},
\end{equation}

and
\begin{equation}
    \ell_A = \pi \frac{D_C(z_\ast)}{r_s(z_\ast)},
\end{equation}

where $D_C(z_\ast)$ is the comoving distance to the photon-decoupling redshift $z_\ast$, and $r_s(z_\ast)$ is the sound horizon evaluated at that redshift. The corresponding chi-square is written as
\begin{equation}
    \chi^2_{\rm CMB}
    =
    \Delta \mathbf{v}_{\rm CMB}^{T}
    \mathbf{C}_{\rm CMB}^{-1}
    \Delta \mathbf{v}_{\rm CMB},
\end{equation}

where
\begin{equation}
    \Delta \mathbf{v}_{\rm CMB}
    =
    \mathbf{v}_{\rm CMB}^{\rm th}
    -
    \mathbf{v}_{\rm CMB}^{\rm obs},
\end{equation}

and $\mathbf{C}_{\rm CMB}$ is the covariance matrix of the compressed CMB observables.

We consider four combinations of the data sets in our analysis. Assuming that the corresponding likelihoods are independent, the total likelihood for each case is given by
\begin{align}
\mathcal{L}_{\rm tot}^{(I)} &\propto \mathcal{L}_{\rm CMB}\,\mathcal{L}_{\rm SN}\, \nonumber \\
\mathcal{L}_{\rm tot}^{(II)} &\propto \mathcal{L}_{\rm CMB}\,\mathcal{L}_{\rm SN}\,\mathcal{L}_{\rm SH0ES}, \nonumber \\
\mathcal{L}_{\rm tot}^{(III)} &\propto \mathcal{L}_{\rm CMB}\,\mathcal{L}_{\rm BAO}\,\mathcal{L}_{\rm SN}, \nonumber\\
\mathcal{L}_{\rm tot}^{(IV)} &\propto \mathcal{L}_{\rm CMB}\,\mathcal{L}_{\rm BAO}\,\mathcal{L}_{\rm SN}\,\mathcal{L}_{\rm SH0ES}, 
\end{align}

where we have adopted the standard relation $\mathcal{L} \propto \exp\left(-\frac{1}{2}\chi^2\right)$ between the likelihood and the chi-square. 
%\begin{align}
%\text{(i)}\quad
%\chi^2_{\rm tot} &= \chi^2_{\rm CMB} + \chi^2_{\rm BAO} + \chi^2_{\rm SN}, \\
%\text{(ii)}\quad
%\chi^2_{\rm tot} &= \chi^2_{\rm CMB} + \chi^2_{\rm BAO} + \chi^2_{\rm SN} + \chi^2_{\rm SH0ES}, \\
%\text{(iii)}\quad
%\chi^2_{\rm tot} &= \chi^2_{\rm CMB} + \chi^2_{\rm SN} + \chi^2_{\rm SH0ES}.
%\end{align}

%Using the standard relation between the likelihood and the chi-square,
%\begin{equation}
%\mathcal{L} \propto \exp\left(-\frac{1}{2}\chi^2\right),
%\end{equation}
%For each data set the total likelihood can be written as

To explore the multidimensional parameter space and reconstruct the posterior distributions of the model parameters, we employ the affine-invariant Markov Chain Monte Carlo (MCMC) sampler implemented in the publicly available \texttt{emcee} package \cite{Foreman_Mackey_2013}. Sampling is performed using the likelihood functions defined above, and the best-fit parameter values are obtained via maximum-likelihood estimation, corresponding to the minimum total chi-square. The resulting chains are subsequently analyzed using the \texttt{GetDist} package \cite{Lewis:2019xzd}, from which we extract the marginalized constraints and construct the corresponding confidence regions at the $1\sigma$, $2\sigma$, and $3\sigma$ levels.

The priors adopted for the free parameters are summarized in Table~\ref{tab:priors}. For the baseline cosmological parameters, we impose broad uniform priors that cover the physically relevant region of parameter space without unduly constraining the posterior distributions. The same approach is applied to the DE parameters in the CPL parametrization, as well as to the EDE amplitude $f_{\rm EDE}$ and the post-critical EoS parameter $w_f$.

For the critical redshift of the EDE component, we impose a Gaussian prior on $\log_{10}(z_c)$, $\log_{10}(z_c) \sim \mathcal{N}(3.544,0.1)$. This choice centers the transition around $z_c \simeq 10^{3.544} \simeq 3500$, corresponding to the epoch close to matter-radiation equality, where an EDE contribution can most efficiently modify the pre-recombination expansion history and the sound horizon. The finite width of the prior allows $z_c$ to vary around this theoretically motivated epoch, while avoiding regions in which the EDE component would peak either too early or too late to have the intended impact on the CMB distance-prior constraints.

To compare the performance of different cosmological models, we use the spatially flat $\Lambda$CDM model as the reference model. We then compare it with three extensions: EDE-$\Lambda$CDM, CPL, and EDE-CPL. The model comparison is performed using the Akaike Information Criterion (AIC) \cite{1100705}, defined as
\begin{equation}
    {\rm AIC} = \chi^2_{\rm min} + 2N,
\end{equation}

where $\chi^2_{\rm min}$ is the minimum chi-square value and $N$ is the number of free parameters of the model. The AIC penalizes models with a larger number of free parameters, and therefore provides a balance between goodness of fit and model complexity. Smaller values of AIC indicate a statistically preferred model.

For each extended model, we compute the difference with respect to the reference flat $\Lambda$CDM model,
\begin{equation}
    \Delta {\rm AIC} = {\rm AIC}_{\rm model} - {\rm AIC}_{\rm reference~ model}.
\end{equation}

Thus, $\Delta{\rm AIC}<0$ indicates that the improvement in the fit is large enough to compensate for the additional parameters, favoring the extended model over the base model. Conversely, $\Delta{\rm AIC}>0$ indicates that the reference model is preferred after penalizing the extra model complexity. As a rule of thumb, $|\Delta{\rm AIC}|<2$ corresponds to no significant preference, $2\lesssim|\Delta{\rm AIC}|\lesssim6$ indicates positive but moderate support, $6\lesssim|\Delta{\rm AIC}|\lesssim10$ indicates strong support, and $|\Delta{\rm AIC}|\gtrsim10$ indicates very strong support for the model with the smaller AIC.

In addition, we quantify the statistical significance of the improvement relative to the reference model using the change in chi-square. Assuming that the extended model is effectively compared to the reference model through $k$ additional degrees of freedom, the probability associated with the improvement is estimated as
\begin{equation}
    p = 1 - F_{\chi^2_k} \left(|\Delta\chi^2|\right),
\end{equation}

where $F_{\chi^2_k}$ is the cumulative distribution function of a chi-square distribution with $k$ degrees of freedom. The corresponding Gaussian-equivalent significance is then obtained from
\begin{equation}
    \sigma = \Phi^{-1} \left(1-\frac{p}{2}\right),
\end{equation}

where $\Phi^{-1}$ is the inverse cumulative distribution function of the standard normal distribution. In practice, this provides an approximate measure of how significant the reduction in $\chi^2$ is, once the number of additional parameters is taken into account \cite{Wilks:1938dza,10.1093/ptep/ptac097,Cowan:2010js}. In the next section, we present our numerical results.

\begin{table}[ht]
\centering
\begin{tabular}{lc}
\hline\hline
Parameter & Prior \\
\hline\hline
$\Omega_b h^2$      & $\mathcal{U}[0.005, 0.05]$ \\
$\Omega_m h^2$      & $\mathcal{U}[0.05, 0.3]$ \\
$h$                 & $\mathcal{U}[0.1, 1.5]$ \\
$n_s$               & $\mathcal{U}[0.8, 1.1]$ \\
$M$                 & $\mathcal{U}[-21.0, -18.0]$ \\
\hline
$w_0$               & $\mathcal{U}[-3.0, 1.0]$ \\
$w_a$               & $\mathcal{U}[-3.0, 2.0]$ \\
\hline
$f_{\rm EDE}$       & $\mathcal{U}[0.0, 0.5]$ \\
$\log_{10}(z_c)$    & $\mathcal{N}(3.544, 0.1)$ \\
$w_f$               & $\mathcal{U}[0.0, 1.0]$ \\
\hline\hline
\end{tabular}
\caption{Priors imposed on the cosmological and EDE parameters used in the MCMC analysis. $\mathcal{U}$ and $\mathcal{N}$ denote Uniform and Normal distributions, respectively.}
\label{tab:priors}
\end{table}

\begin{table*}[ht]
\centering
\footnotesize
\renewcommand{\arraystretch}{1.5}
\begin{tabular}{lccccccccc}
\hline\hline
Model & $H_0$ & $\Omega_m$ & $M$ & $w_0$ & $w_a$ & $f_{\rm EDE}$ & $\log_{10}(z_c)$ & $w_f$ & $\chi^2_{min}$ \\
\hline\hline
$\Lambda$CDM 
& $67.2\pm 0.5$
& $0.318\pm 0.007$
& $-19.45\pm 0.01$
& --
& --
& --
& --
& --
& $1457.9$
\\

EDE-$\Lambda$CDM 
& $74.4^{+3.3}_{-5.6}$
& $0.319^{+0.007}_{-0.009}$
& $-19.23^{+0.10}_{-0.16}$
& --
& -
& $0.162^{+0.069}_{-0.120}$
& $3.53\pm 0.10$
& $0.64^{+0.30}_{-0.34}$
& $1457.5$
\\
\hline

CPL
& $67.3\pm 1.3$
& $0.316^{+0.011}_{-0.013}$
& $-19.43^{+0.05}_{-0.04}$
& $-0.90\pm 0.10$
& $-0.41^{+0.56}_{-0.48}$
& --
& --
& --
& $1456.6$
\\

EDE-CPL
& $74.4^{+3.6}_{-6.5}$
& $0.319^{+0.011}_{-0.016}$
& $-19.22^{+0.11}_{-0.19}$
& $-0.91\pm 0.11$
& $-0.32^{+0.62}_{-0.49}$
& $0.165^{+0.070}_{-0.13}$
& $3.55^{+0.12}_{-0.090}$
& $0.63^{+0.29}_{-0.18}$
& $1456.8$
\\
\hline\hline
\end{tabular}
\caption{Cosmological Constraints from Joint Analysis of CMB and Pantheon+ (without SH0ES and in the absence of DESI BAO)}
\label{tab:cosmo_models1}
\end{table*}

\begin{table}[h!]
\centering
\begin{tabular}{lccccc}
\hline\hline
Model & $\chi^2_{min}$ & $\Delta\chi^2_{min}$ & $\Delta$AIC & Deviation \\
\hline
$\Lambda$CDM & $1457.9$ & $0.0$ & $0.0$ & -- \\
EDE-$\Lambda$CDM & $1457.5$ & $-0.4$ & $+5.6$ & $0.1\sigma$ \\
CPL & $1456.6$ & $-2.3$ & $+1.7$ & $1.0\sigma$ \\
EDE-CPL & $1456.8$ & $-2.1$ & $+7.9$ & $0.2\sigma$ \\
\hline\hline
\end{tabular}
\caption{The values of $\chi^2_{min}$, $\Delta \chi^2_{min}$, $\Delta \text{AIC}$ and deviation in different models, computed using the combined CMB and Pantheon+ (without SH0ES and in the absence of DESI BAO)}
\label{tab:chi2_1}
\end{table}
%========================================================
 \section{Numerical results}\label{sect:num_results}
 We now present the numerical results of our MCMC analysis, which we divide into two distinct cases. The first case consists of a joint analysis of CMB and Pantheon+ without DESI BAO, whereas the second case includes DESI BAO. For each case, we further perform the analysis both with and without the SH0ES SNeIa data. The comparison with and without DESI BAO serves to determine whether the inclusion of such data induces a deviation from the standard cosmological model. Similarly, the comparison with and without SH0ES SNeIa elucidates the effect of the Cepheid calibration of SNeIa on our analysis.
 
 \subsection{A joint analysis without DESI BAO}
\begin{figure}[htbp]
    \centering
    \includegraphics[width=0.5\textwidth]{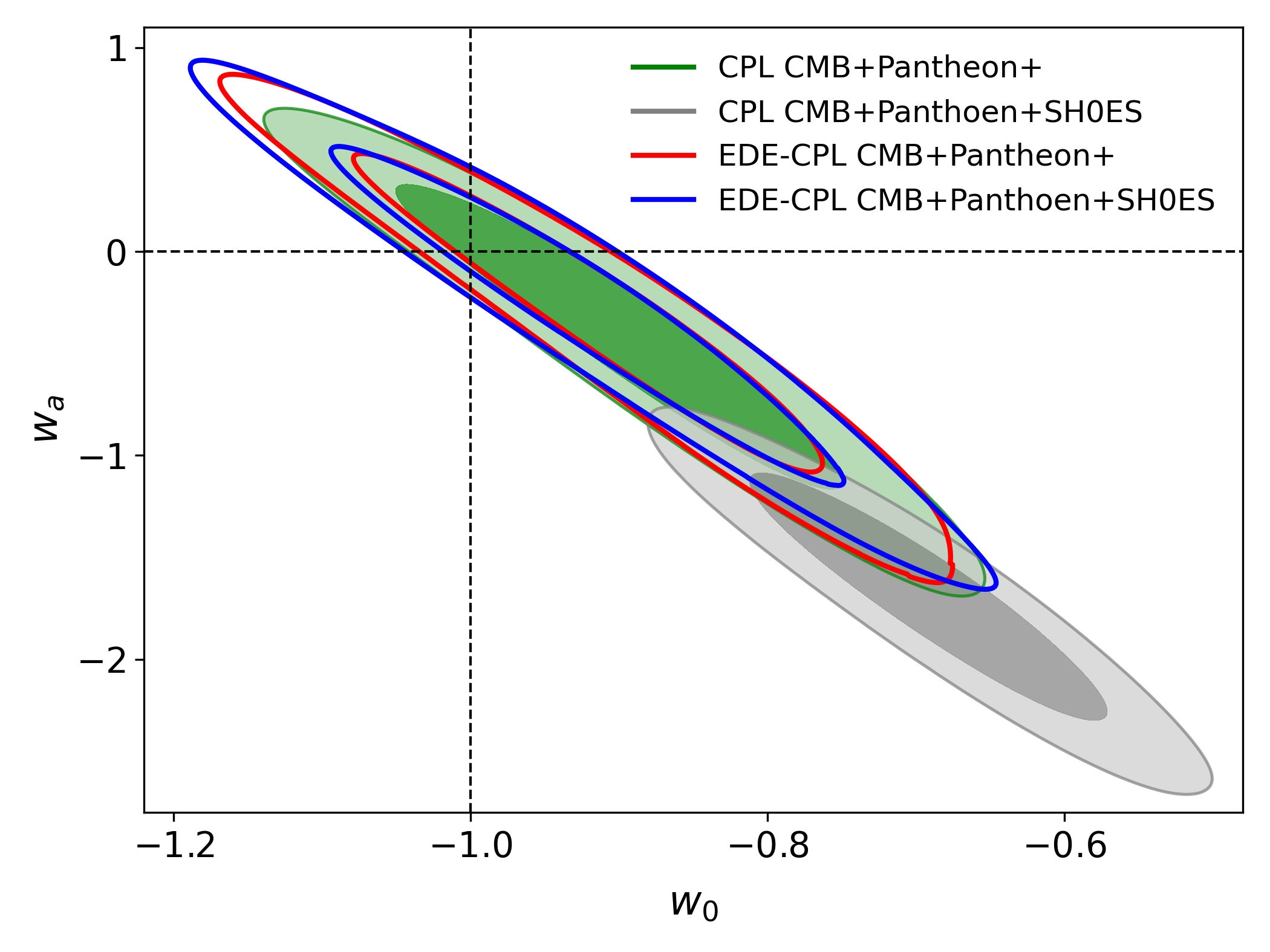}
    \caption{Two-dimensional marginalized posterior distributions for the EoS parameters, $w_0$ and $w_a$. The inner and outer contours represent the 68\% ($1\sigma$) and 95\% ($2\sigma$) confidence levels, respectively, for the various cosmological models and dataset combinations considered in this work.}
    \label{fig:fig1}
\end{figure}

 \noindent \textbf{CMB+Pantheon+ (without SH0ES)}:\\
 The numerical results, including the observational constraints on the cosmological parameters and comparison analysis, for different cosmological models using the joint CMB+Pantheon+ (without the SH0ES SNeIa data), are respectively presented in Tables \ref{tab:cosmo_models1} and \ref{tab:chi2_1}. Without EDE, we see that the value of the Hubble constant $H_0$ is close to the Planck value for both $\Lambda$CDM and CPL models. It should be noted that, in the absence of the SH0ES data, the joint analysis of the CMB and uncalibrated SNeIa (Pantheon+) is strongly dominated by the CMB. Since the CMB has a significantly greater effect on our statistical constraints than uncalibrated Pantheon+, the posterior distributions are primarily driven by the CMB data. Consequently, for the standard $\Lambda$CDM cosmology and CPL case, the constrained value of the Hubble constant $H_0$ remains firmly bounded within the standard Planck range $H_0 = 67.36 \pm 0.54 \text{ km s}^{-1} \text{ Mpc}^{-1}$ \cite{Planck:2018vyg}. Due to the intrinsic degeneracy between $H_0$ and the absolute magnitude $M$, this tight constraint on the Hubble constant correspondingly anchors the absolute magnitude at $M \approx -19.45$ for these models. Furthermore, for the joint CMB and Pantheon+ dataset without the SH0ES data, the constrained values for EoS parameters of CPL are well consistent with the standard $\Lambda$CDM paradigm ($w_0 = -1$ and $w_a = 0$) within the $1\sigma$ confidence level (see also the green contour in Fig.~\ref{fig:fig1}). From a statistical perspective, the CPL parametrization yields a minimum chi-square value compatible with $\Lambda$CDM value. This means that CPL does not indicate any statistical deviation from the standard cosmological model and, consequently, there is no hint of an evolving DE scenario in the presence of joint CMB and Pantheon+ data. Turning our attention to the cosmological models invoking EDE, EDE-$\Lambda$CDM and EDE-CPL, we observe a significant shift in the Hubble constant $H_0$ and absolute magnitude $M$, as expected. The inclusion of the EDE component effectively reduces the sound horizon at the epoch of recombination, and to maintain the observed angular positions of the CMB acoustic peaks, this reduction naturally drives the Hubble constant to higher values in both EDE-based models, which successfully alleviates the Hubble tension. Concurrently, due to the tight $H_0-M$ degeneracy, the preferred value of the absolute magnitude is very close to the local SH0ES prior $M = -19.253 \pm 0.027 \text{ mag}$ \cite{Riess:2021jrx} for both EDE-extended models, despite the explicit absence of the SH0ES data in our data analysis. This clearly highlights the intrinsic capability of the EDE model to simultaneously adjust both $H_0$ and $M$ towards their local measurements, as expected. We also observe that for both EDE-$\Lambda$CDM and EDE-CPL models, the EDE parameters, the maximum fractional contribution of EDE, $f_{\rm EDE}$, the critical redshift of the transition, $z_c$, and the EoS parameter of EDE, $w_f$ obtained in our analysis are consistent with  previous studies investigating the efficacy of EDE in resolving the Hubble tension \cite{Poulin:2018cxd,Chaussidon:2025npr}. Indeed, an EDE injection of $f_{\rm EDE} \approx 0.1 - 0.15$ is typically required to fully bridge the gap between early-time CMB measurements and local distance-ladder calibrations. The constrained value of $w_f$ well above the EoS of radiation ($w_r=1/3$) indicates that following the phase transition at $z_c$, the EDE fluid dilutes rapidly. This fast dilution is a crucial theoretical requirement for successful EDE models, as it ensures that the extra energy component decays quickly enough to avoid spoiling the highly constrained evolution of density perturbations and the resulting CMB power spectra at later times.
 Furthermore, a closer inspection of the EDE-CPL framework  reveals that both constrained values of the EoS parameters $w_0$ and $w_a$ remain entirely consistent with the standard cosmological constant points within the $1\sigma$ confidence level (see also the red contour in Fig.~\ref{fig:fig1}), reiterating the lack of substantial evidence for late-time DE. In addition, the minor differences $\Delta \chi^2$ between the EDE-extended models and the CPL model with a standard $\Lambda$CDM cosmology indicate that neither late-time evolving DE nor EDE provides a significantly superior statistical fit to the combined CMB and Pantheon+ dataset.
 It is notable that while the inferred EDE parameters in our analysis (without explicit SH0ES data) naturally prefer values that resolve the $H_0$ discrepancy, the overall statistical fit---as indicated by the AIC---remains penalized relative to the standard $\Lambda$CDM model due to the introduction of additional parameters. Furthermore, the similarity of the constrained values of EDE parameters and also the similarity of $\chi^2$ values in both EDE-$\Lambda$CDM and EDE-CPL models ($AIC$ of EDE-CPL is $2.3$ larger than EDE-$\Lambda$CDM) reinforces the conclusion that modifications to the late-time expansion history are largely subdominant when a robust EDE mechanism is present. Notice that this conclusion is obtained without  invoking both the SH0ES data and DESI BAO measurements.\\
%========================================================

\begin{table*}[t]
\centering
\footnotesize
\renewcommand{\arraystretch}{1.5}
\begin{tabular}{lccccccccc}
\hline\hline
Model & $H_0$ & $\Omega_m$ & $M$ & $w_0$ & $w_a$ & $f_{\rm EDE}$ & $\log_{10}(z_c)$ & $w_f$ & $\chi^2_{min}$ \\
\hline\hline
$\Lambda$CDM 
& $68.4\pm 0.5$
& $0.302\pm 0.006$
& $-19.41 \pm 0.01$
& --
& --
& --
& --
& --
& $1550.3$
\\

EDE-$\Lambda$CDM 
& $73.5\pm 1.1$
& $0.319^{+0.006}_{-0.009}$
& $-19.25\pm 0.03$
& --
& --
& $0.147^{+0.031}_{-0.043}$
& $3.55\pm 0.09$
& $0.61\pm 0.23$
& $1514.8$
\\
\hline

CPL
& $70.6\pm 0.8$
& $0.286\pm 0.007$
& $-19.32\pm 0.02$
& $-0.69\pm 0.08$
& $-1.69\pm 0.40$
& --
& --
& --
& $1528.3$
\\

EDE-CPL
& $73.4\pm 1.1$
& $0.319^{+0.012}_{-0.015}$
& $-19.25\pm 0.03$
& $-0.92\pm 0.12$
& $-0.30\pm 0.56$
& $0.158^{+0.044}_{-0.054}$
& $3.55^{+0.12}_{-0.10}$
& $0.67^{+0.27}_{-0.15}$
& $1514.0$
\\
\hline\hline
\end{tabular}
\caption{Cosmological Constraints from Joint Analysis of CMB and Pantheon+SH0ES dataset (the absence of DESI BAO)}
\label{tab:cosmo_models2}
\end{table*}

\begin{table}
\centering
\begin{tabular}{lccccc}
\hline\hline
Model & $\chi^2_{min}$ & $\Delta\chi^2_{min}$ & $\Delta$AIC & Deviation \\
\hline
$\Lambda$CDM & $1550.3$ & $0.0$ & $0.0$ & -- \\
EDE-$\Lambda$CDM & $1514.8$ & $-35.5$ & $-29.5$ & $5.3\sigma$ \\
CPL & $1528.3$ & $-22.0$ & $-18.0$ & $4.3\sigma$ \\
EDE-CPL & $1514.0$ & $-36.3$ & $-26.3$ & $4.9\sigma$ \\
\hline\hline
\end{tabular}
\caption{The values of $\chi^2_{min}$, $\Delta \chi^2_{min}$, $\Delta \text{AIC}$, and deviations in different models, computed using the CMB and Pantheon+SH0ES (the absence of DESI BAO)}
\label{tab:chi2_2}
\end{table}

%=
\begin{table*}[t]
\centering
\footnotesize
\renewcommand{\arraystretch}{1.5}
\begin{tabular}{lccccccccc}
\hline\hline
Model & $H_0$ & $\Omega_m$ & $M$ & $w_0$ & $w_a$ & $f_{\rm EDE}$ & $\log_{10}(z_c)$ & $w_f$ & $\chi^2_{min}$ \\
\hline\hline
$\Lambda$CDM 
& $68.3 \pm 0.3$
& $0.303 \pm 0.004$
& $-19.42 \pm 0.01$
& --
& --
& --
& --
& --
& $1475.7$
\\

EDE-$\Lambda$CDM 
& $84.0^{+10.0}_{-6.0}$
& $0.305 \pm 0.004$
& $-18.98^{+0.27}_{-0.13}$
& --
& --
& $0.314^{+0.160}_{-0.070}$
& $3.56 \pm 0.09$
& $0.77^{+0.18}_{-0.11}$
& $1471.2$
\\
\hline

CPL
& $67.6 \pm 0.6$
& $0.312 \pm 0.006$
& $-19.42 \pm 0.01$
& $-0.84^{+0.05}_{-0.06}$
& $-0.58^{+0.23}_{-0.19}$
& --
& --
& --
& $1467.2$
\\

EDE-CPL 
& $79.7 \pm 6.3$
& $0.313 \pm 0.006$
& $-19.07^{+0.20}_{-0.17}$
& $-0.85 \pm 0.06$
& $-0.53 \pm 0.21$
& $0.262^{+0.150}_{-0.086}$
& $3.54 \pm 0.10$
& $0.73^{+0.21}_{-0.13}$
& $1466.2$
\\
\hline\hline
\end{tabular}
\caption{Cosmological Constraints from Joint Analysis of CMB, DESI BAO (DR2), and Pantheon+ (without SH0ES) datasets.}
\label{tab:cosmo_models3}
\end{table*}
%========================================================

\begin{table}
\centering
\begin{tabular}{lccccc}
\hline\hline
Model & $\chi^2_{min}$ & $\Delta\chi^2_{min}$ & $\Delta$AIC & Deviation \\
\hline
$\Lambda$CDM & $1475.7$ & $0.0$ & $0.0$ & -- \\
EDE-$\Lambda$CDM & $1471.2$ & $-4.5$ & $+1.5$ & $1.2\sigma$ \\
CPL & $1467.2$ & $-8.5$ & $-4.5$ & $2.5\sigma$ \\
EDE-CPL & $1466.2$ & $-9.5$ & $+0.5$ & $1.7\sigma$ \\
\hline\hline
\end{tabular}
\caption{The values of $\chi^2_{min}$, $\Delta \chi^2_{min}$, $\Delta \text{AIC}$ and deviation in different models, computed using the combined CMB, DESI BAO (DR2) and Pantheon+ (without SH0ES) dataset.}
\label{tab:chi2_3}
\end{table}
%========================================================
\noindent \textbf{CMB + Pantheon+ SH0ES}:\\
We now switch our focus to the combination of CMB + Pantheon+ SH0ES SNeIa, as detailed in Tables \ref{tab:cosmo_models2} and \ref{tab:chi2_2}. The inclusion of the SH0ES data introduces a critical test for the cosmological models, assessing their ability to reconcile early-universe CMB constraints with the local $H_0$ measurements.
For the standard $\Lambda$CDM model, despite the inclusion of the SH0ES data, the Hubble constant is anchored at $H_0 = 68.4 \pm 0.5 \text{ km/s/Mpc}$. This result shows that while the Cepheid-calibrated SNeIa within the Pantheon+ compilation inherently favor a higher, local $H_0$ value, the large statistical weight of the CMB observations for the rigid framework of standard $\Lambda$CDM cosmology firmly holds $H_0$ down. Consequently, the standard $\Lambda$CDM cosmology struggles to accommodate the SH0ES data, leading to a poor fit to the combined data. Conversely, in the case of CPL parametrization, the additional degrees of freedom governing the late-time DDE, $w_0$ and $w_a$, provide the necessary flexibility to alleviate the tension.  In the CPL case, we observe that the Hubble constant shifts upward to $H_0 = 70.6 \pm 0.8 \text{ km/s/Mpc}$. 
Most notably, the parameters of the EoS of DE, $w_0$ and $w_a$, deviate substantially from their standard cosmological constant values ($w_0=-1, w_a=0$) as reported in Table \ref{tab:cosmo_models2} (see also the gray in Fig.~\ref{fig:fig1}). This translates to a roughly $4.3\sigma$ deviation from the standard $\Lambda$CDM model, with the CPL parameterization yielding a substantially improved fit to the combined dataset ($\Delta \chi^2_{min}=-22.0$). Consequently, the inclusion of the SH0ES measurements provides compelling evidence for a modified late-time expansion history driven by a DDE scenario \cite[see also][]{Scolnic:2021amr}.
In fact, the physical origin of this finding lies in the constraining power of the datasets we have used and the rigidity of the cosmological models under study. In standard $\Lambda$CDM cosmology, the minimal set of free parameters is tightly constrained by the high-precision CMB data, leaving no room to successfully fit the local SH0ES measurements. However, in the CPL parametrization, we have the extra $w_0$ and $w_a$ parameters, which operate dynamically at late times and can influence the late-time expansion history of the Universe to better accommodate the SH0ES data while maintaining consistency with the CMB. 
This improvement is quantitatively supported by the goodness-of-fit statistics reported in Table \ref{tab:chi2_2}. The CPL parametrization yields a drastically improved fit compared to standard $\Lambda$CDM, with a $\Delta AIC = -18.0$ indicating a strong preference for DDE scenarios and ruling out the standard $\Lambda$CDM cosmology.\\
We now turn our attention to cosmological models that invoke EDE. As reported in Table \ref{tab:cosmo_models2}, for both EDE-$\Lambda$CDM and EDE-CPL, the presence of EDE successfully drives the Hubble constant to higher values, matching the local SH0ES measurements $H_0 = 73.04 \pm 1.04 \text{ km s}^{-1} \text{ Mpc}^{-1}$ \cite{Riess:2021jrx}. We observe a huge reduction in the $\chi^2$ values upon the inclusion of EDE, suggesting that the cosmological parameters constrained by the CMB in the EDE framework yield a high level of goodness-of-fit to the SH0ES SNeIa data. This level of consistency  is not achievable with standard $\Lambda$CDM cosmology or the CPL parametrization. Specifically, Table \ref{tab:chi2_2} reveals that EDE-$\Lambda$CDM is statistically preferred over standard $\Lambda$CDM with a profound $\Delta\chi^2_{\min} = -35.5$ ($\Delta \text{AIC} = -29.5$), corresponding to a roughly $5.3\sigma$ significance. Furthermore, the EDE-$\Lambda$CDM model is statistically preferred over the CPL parametrization, as indicated by a difference in AIC of $\Delta \text{AIC} \approx -11.5$. This value strongly implies that, in the presence of the SH0ES data, modifying the early-universe expansion history via EDE provides a more powerful physical solution than introducing DDE at the late universe.
Crucially, examining the EDE-CPL model—which simultaneously allows for both early-time (EDE) and late-time (DDE) modifications—reveals a profound physical signal favored by the combined CMB+Pantheon+SH0ES data. A comparison of the constrained parameters reported in Table \ref{tab:cosmo_models2} shows that EDE-CPL offers no tangible statistical advantage over EDE-$\Lambda$CDM. Both models predict nearly identical values for $H_0$, and their $\chi^2_{\min}$ values are statistically indistinguishable. Hence, the addition of the two late-time parameters $w_0$ and $w_a$ fails to meaningfully improve the goodness-of-fit.
Interestingly, the late-time EoS parameters in the EDE-CPL model are constrained to $w_0 = -0.92 \pm 0.12$ and $w_a = -0.30 \pm 0.56$. 
Unlike the pure CPL model, which exhibited a $4.3\sigma$ deviation from $\Lambda$CDM, these values are entirely consistent with the standard cosmological constant ($w_0=-1, w_a=0$) well within the $1\sigma$ confidence level (see also the blue contour in Fig.~\ref{fig:fig1}).
This demonstrates that when simultaneously allowing for both early- and late-time modifications to the cosmic expansion history, the combined CMB+Pantheon+SH0ES datasets strongly favor the EDE scenario. Once the EDE model successfully alleviates the $H_0$ tension by reducing the sound horizon, the necessity for late-time DDE is significantly diminished, rendering the CPL parameters statistically consistent with a cosmological constant. Consequently, the joint analysis indicates that deviations from the standard $\Lambda$CDM cosmology are better accommodated by modifications to early-universe physics rather than by late-time modifications to the Hubble parameter at low redshifts. It is worth noting that this result is obtained without including DESI BAO measurements in our dataset. In the next step, we incorporate the DESI BAO data into our statistical analysis to determine whether the DDE signal suggested by DESI observations is robust or can be effectively mimicked by an EDE scenario.

\subsection{A joint analysis with DESI BAO}
\begin{figure}[htbp]
    \centering
    \includegraphics[width=0.5\textwidth]{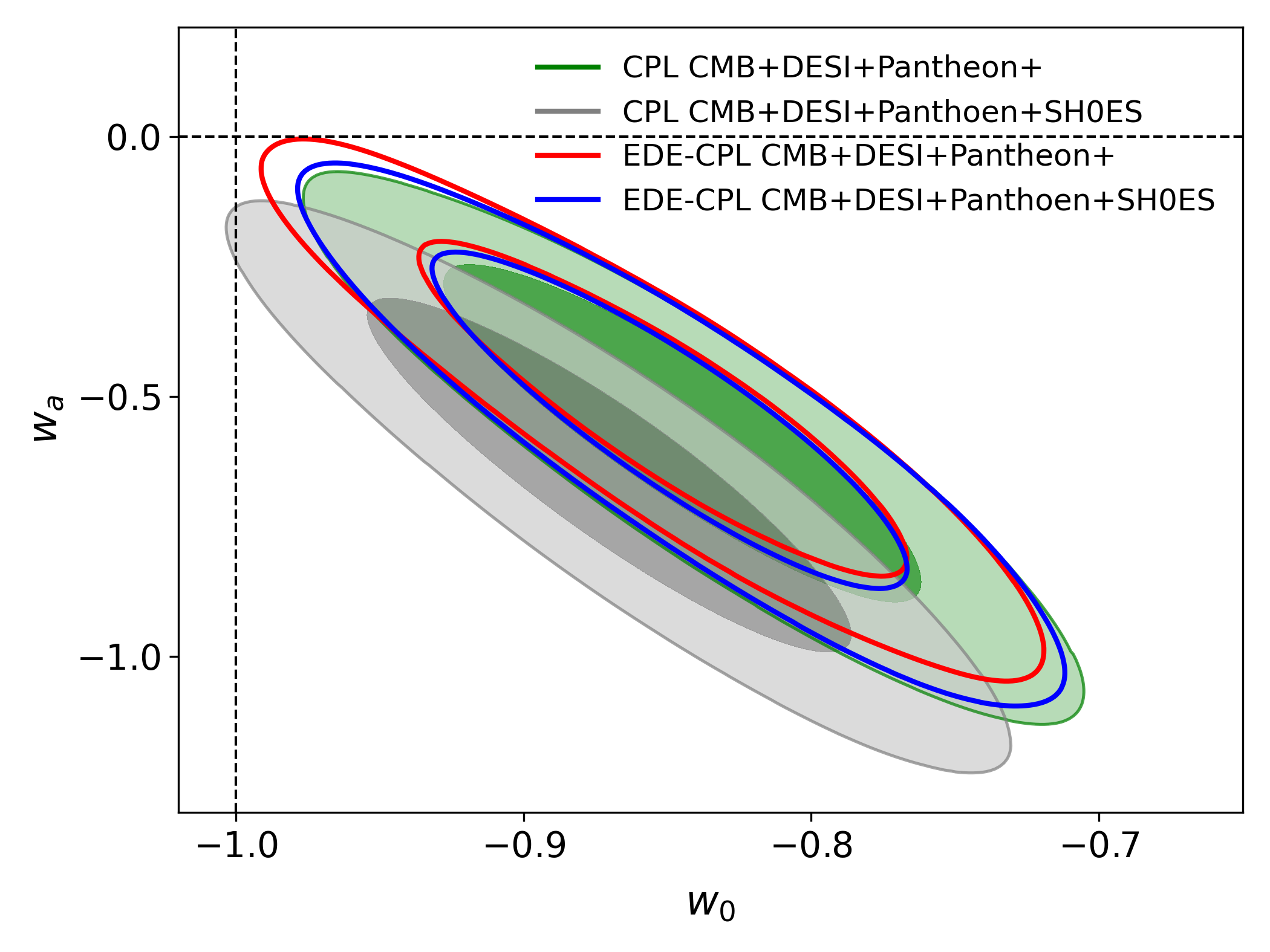}
    \caption{Same as Fig.~\ref{fig:fig1}, but for the dataset combinations incorporating the DESI BAO measurements.}
    \label{fig:fig2}
\end{figure}

\noindent \textbf{CMB + DESI BAO + Pantheon+ (without SH0ES)}:\\
We now investigate the impact of DESI BAO (DR2) data in our analysis, utilizing the dataset combination of CMB + DESI BAO + Pantheon+. The corresponding parameter constraints and statistical model-comparison metrics are presented in Tables \ref{tab:cosmo_models3} and \ref{tab:chi2_3}, respectively.
Starting with the cosmological models without invoking EDE, we observe that the introduction of DESI BAO data drives a strong preference for late-time DDE. One can see that in the CPL parametrization, the EoS parameters are constrained to $w_0 = -0.84^{+0.05}_{-0.06}$ and $w_a = -0.58^{+0.23}_{-0.19}$, showing a clear deviation from the standard $\Lambda$CDM cosmology (see also the green contour in Fig.~\ref{fig:fig2}). We emphasize that this deviation is basically driven by the inclusion of the DESI BAO measurements. In our previous analysis using the same dataset combination without DESI BAO, the CPL parameters remained statistically consistent with the standard $\Lambda$CDM cosmology. This dynamical behavior leads to a significant improvement in the fit, yielding $\Delta\chi^2 = -8.5$ compared to the $\Lambda$CDM model \cite[see also][]{DESI:2025zgx}. Consequently, the CPL parametrization emerges as the statistically preferred case in this dataset combination, with a $\Delta\mathrm{AIC} = -4.5$ and an equivalent deviation of roughly $2.5\sigma$. Notice that in the absence of EDE, the Hubble constant in this case remains anchored at $H_0 = 67.6 \pm 0.6$ km/s/Mpc, reinforcing the conclusion that late-time modifications alone are insufficient to resolve the Hubble tension. 
When EDE is included in our cosmological models, we observe a considerable shift in the parameters. In the EDE-$\Lambda$CDM model, the Hubble constant surges to $H_0 = 84.0^{+10.0}_{-6.0}$ km/s/Mpc, while the absolute magnitude of SNe Ia shifts to $M = -18.98^{+0.27}_{-0.13}$. Notably, in the absence of the SH0ES data, the inclusion of DESI BAO data allows the EDE fraction to become excessively large ($f_{\rm EDE} = 0.314^{+0.160}_{-0.070}$). This unconstrained growth of $f_{\rm EDE}$ artificially inflates $H_0$ well beyond its local measurement and pulls $M$ significantly away from the Cepheid-calibrated value, albeit accompanied by substantially large uncertainties.
Combining both early- and late-time modifications in the EDE-CPL cosmological model yields the lowest overall minimum chi-square ($\chi^2_{\min} = 1466.2$). Interestingly, the CPL parameters in this extended model ($w_0 = -0.85 \pm 0.06$, $w_a = -0.53 \pm 0.21$) remain almost identical to those in the base CPL parametrization (see red contour in Fig.~\ref{fig:fig2}), confirming that the DESI BAO preference for DDE persists independently of the modification to early-time physics. Furthermore, the EDE fraction in EDE-CPL remains large ($f_{\rm EDE} = 0.262^{+0.150}_{-0.086}$), leading to an overestimated Hubble constant of $H_0 = 79.7 \pm 6.3$ km/s/Mpc. 
In the absence of SH0ES data, we observe that according to the AIC, we have $\Delta\mathrm{AIC} = +1.5$ and $\Delta\mathrm{AIC} = +0.5$ for the EDE-$\Lambda$CDM and EDE-CPL cosmological models, respectively, relative to the standard $\Lambda$CDM cosmology. In addition, there are only $1.2\sigma$ and $1.7\sigma$ deviations in improving the fit to the combined data for the EDE-$\Lambda$CDM and EDE-CPL models, respectively. Hence, CPL is the sole statistically preferred extension for the combination of CMB + DESI BAO + Pantheon+ datasets, in agreement with the results of DESI collaboration \cite{DESI:2024mwx,DESI:2025zgx}.
It is important to note that in our previous analysis without SH0ES (where DESI BAO was excluded), we observed that the EoS parameters of the CPL parametrization were well within the bounds of the $\Lambda$CDM cosmology, showing no significant deviation. Therefore, the distinct preference for DDE observed here is driven entirely by the addition of the DESI BAO data to our analysis.
%========================================================
\begin{table*}[t]
\centering
\footnotesize
\renewcommand{\arraystretch}{1.5}
\begin{tabular}{lccccccccc}
\hline\hline
Model & $H_0$ & $\Omega_m$ & $M$ & $w_0$ & $w_a$ & $f_{\rm EDE}$ & $\log_{10}(z_c)$ & $w_f$ & $\chi^2_{min}$ \\
\hline\hline
$\Lambda$CDM
& $68.7 \pm 0.3$
& $0.298 \pm 0.003$
& $-19.40 \pm 0.01$
& --
& --
& --
& --
& --
& $1561.3$
\\

EDE-$\Lambda$CDM
& $73.9 \pm 1.0$
& $0.303 \pm 0.004$
& $-19.25 \pm 0.03$
& --
& --
& $0.150^{+0.033}_{-0.037}$
& $3.56 \pm 0.10$
& $0.76^{+0.20}_{-0.11}$
& $1531.2$
\\
\hline

CPL
& $68.8 \pm 0.6$
& $0.302 \pm 0.005$
& $-19.39 \pm 0.01$
& $-0.87 \pm 0.06$
& $-0.66^{+0.24}_{-0.21}$
& --
& --
& --
& $1550.9$
\\

EDE-CPL
& $73.1 \pm 1.1$
& $0.312 \pm 0.006$
& $-19.25 \pm 0.03$
& $-0.85 \pm 0.06$
& $-0.56^{+0.23}_{-0.20}$
& $0.143 \pm 0.036$
& $3.55 \pm 0.10$
& $0.71^{+0.23}_{-0.15}$
& $1524.1$
\\
\hline\hline
\end{tabular}
\caption{Cosmological Constraints from Joint Analysis of CMB, DESI BAO (DR2) and Pantheon+SH0ES data.}
\label{tab:cosmo_models4}
\end{table*}
%========================================================

\begin{table}
\centering
\begin{tabular}{lccccc}
\hline\hline
Model & $\chi^2_{min}$ & $\Delta\chi^2_{min}$ & $\Delta$AIC & Deviation \\
\hline
$\Lambda$CDM & $1561.3$ & $0.0$ & $0.0$ & -- \\
EDE-$\Lambda$CDM & $1531.2$ & $-30.1$ & $-24.1$ & $4.8\sigma$ \\
CPL & $1550.9$ & $-10.4$ & $-6.4$ & $2.8\sigma$ \\
EDE-CPL & $1524.1$ & $-37.2$ & $-27.2$ & $5.0\sigma$ \\
\hline\hline
\end{tabular}
\caption{The values of $\chi^2_{min}$, $\Delta \chi^2_{min}$, $\Delta \text{AIC}$ and deviation in different models, computed using the combined CMB, DESI BAO (DR2) and Pantheon+SH0ES.}
\label{tab:chi2_4}
\end{table}
\noindent \textbf{CMB + DESI BAO + Pantheon+ SH0ES}:\\
Finally, we incorporate the SH0ES data alongside the CMB, DESI BAO, and Pantheon+ datasets. The numerical constraints for the cosmological parameters are summarized in Table~\ref{tab:cosmo_models4}, and the corresponding statistical comparisons are presented in Table~\ref{tab:chi2_4}. Starting with the base models, we observe that for both $\Lambda$CDM and CPL, the inferred values of the Hubble constant remain heavily tethered to the Planck value due to the overwhelming statistical weight of the CMB data compared to the other datasets. Furthermore, the inclusion of the SH0ES prior introduces a strong internal tension within this dataset combination. As expected, the CPL parametrization attempts to accommodate this discrepancy by exhibiting a larger $2.8\sigma$ deviation from $\Lambda$CDM ($\Delta\text{AIC} = -6.4$) compared to the case without SH0ES (Table~\ref{tab:chi2_3}), with its EoS parameters $w_0$ and $w_a$ correspondingly shifting away from the constant $\Lambda$CDM cosmology $w_0=-1, w_a=0$ (see also gray contour in Fig.~\ref{fig:fig2}). Turning to the EDE-based models, which successfully resolve the $H_0$ tension, we observe that both the EDE-$\Lambda$CDM and EDE-CPL models yield $H_0 \sim 73 \pm 1 \text{ km s}^{-1} \text{Mpc}^{-1}$ and provide the best overall fits. Specifically, EDE-$\Lambda$CDM yields $\chi^2_{\min} = 1531.2$ ($\Delta\text{AIC} = -24.1$), while EDE-CPL reaches $\chi^2_{\min} = 1524.1$ ($\Delta\text{AIC} = -27.2$). 
Comparing these results with the analysis lacking the SH0ES data (Tables~\ref{tab:cosmo_models3} and \ref{tab:chi2_3}) clearly illustrates the crucial stabilizing role of the local Cepheid calibration, as well as its profound impact on model selection. Statistically, the SH0ES data strongly penalizes the low $H_0$ prediction of the $\Lambda$CDM model, forcing the extended models to diverge further from the standard baseline to absorb this tension. Consequently, the statistical deviations from the standard model systematically increase across all extensions; for instance, the CPL deviation grows from $2.5\sigma$ (in the absence of SH0ES) to $2.8\sigma$, and the EDE-based models are driven to highly significant deviations of $4.8\sigma$ and $5.0\sigma$. In particular, the departure of the EDE-CPL model from the standard cosmology is substantially more robust here compared to its marginal deviation in the scenario without the local prior. 
Subsequently, comparing the full analysis with the case excluding DESI BAO (Tables~\ref{tab:cosmo_models2} and \ref{tab:chi2_2}) reveals a critical shift in model preference driven by the DESI data. When DESI BAO was excluded, EDE-$\Lambda$CDM was statistically favored, and the EoS parameters $w_0$ and $w_a$ in the EDE-CPL model remained consistent with a cosmological constant ($w_{\Lambda}=-1.0$) within the $1\sigma$ confidence level. In contrast, the inclusion of the DESI data induces a significant departure of $w_0$ and $w_a$ from the $\Lambda$CDM baseline (see the blue contour in Fig.~\ref{fig:fig2}). In addition, we observe a moderate statistical preference for the extended EDE-CPL model ($|\Delta\text{AIC}| = 3.1$) relative to EDE-$\Lambda$CDM. Hence, while EDE provides the necessary early-time modification to resolve the Hubble tension, our full dataset analysis suggests that a late-time DDE modification is still required to describe the DESI BAO observations comprehensively

 \section{Conclusions}\label{sect:conclusion}
In this work, our primary objective was to evaluate whether EDE can serve as a viable alternative to the late-time DDE recently signaled by DESI BAO observations. While it is well established in the literature that EDE provides a robust mechanism for alleviating the Hubble tension, its potential interplay with recent hints of late-time dynamics requires careful investigation. By employing combinations of Planck CMB data, Pantheon+ SNeIa, the latest DESI BAO (DR2) measurements, and the Cepheid-calibrated SNeIa (SH0ES data), we performed a rigorous comparative analysis of four distinct models: $\Lambda$CDM, CPL, EDE-$\Lambda$CDM, and EDE-CPL. Ultimately, we sought to determine whether modifying the pre-recombination sound horizon via EDE absorbs or diminishes the statistical preference for DDE, or if the current DESI BAO cosmological data genuinely demand the evolving DE scenario at late times. Through a progressive analysis of different dataset combinations, we draw the following conclusions:\\
\begin{itemize}
\item First, focusing on data combinations without DESI BAO measurements—both with and without the local SH0ES calibration—we found that in the absence of SH0ES data, there is no statistical preference for an evolving dark energy (DE) component in the late Universe. Specifically, the equation-of-state (EoS) parameters in the CPL parameterization ($w_0$ and $w_a$) remain entirely consistent with a cosmological constant. This behavior is naturally extended to EDE cosmologies, where the late-time expansion in the EDE-CPL model exhibits full agreement with the $\Lambda$CDM background, showing no significant deviation in its EoS parameters. Consequently, none of the extended models (CPL, EDE-$\Lambda$CDM, or EDE-CPL) yield a statistically superior fit to this data combination compared to the standard $\Lambda$CDM model. In these scenarios, the EDE component fulfills its primary theoretical role: it reduces the sound horizon at the recombination epoch to accommodate a higher $H_0$, thereby addressing the Hubble tension without the need for additional late-time dynamical features. Conversely, when incorporating the local SH0ES calibration, the CPL parametrization acting as a DDE model yields a significantly improved fit compared to the standard $\Lambda$CDM cosmology. In this scenario, the extra degrees of freedom ($w_0$ and $w_a$) deviate notably from the fixed values of a cosmological constant ($w_0=-1.0 \; \& \; w_a=0.0$), thereby accommodating the SH0ES data without compromising the fit to the combined CMB and Pantheon+ observations. Interestingly, we find that the inclusion of an EDE component effectively supplants the need for late-time DDE. Specifically, the EDE-$\Lambda$CDM cosmology provides a statistically superior fit compared to the CPL model. Furthermore, when both extensions are combined in the EDE-CPL model, the EoS parameters $w_0$ and $w_a$ remain entirely consistent with a cosmological constant. Given that the EDE-CPL and EDE-$\Lambda$CDM models provide fits of equal statistical quality, this strongly implies that an early-universe modification (EDE) entirely absorbs the phenomenological preference for late-time DDE dynamics; thus, in the presence of EDE, introducing DDE becomes completely redundant.

\item In the next phase of our analysis, we incorporate the DESI BAO measurements. In the absence of the local SH0ES calibration, this updated data combination exhibits a distinct preference for DDE, manifesting as a noticeable deviation from $\Lambda$CDM within the CPL framework. This departure is entirely driven by the DESI BAO data, given that our baseline analysis excluding DESI showed no such deviation. When introducing the EDE component into this setup, the preference for DDE—indicated by the shift in the CPL parameters—persists. However, lacking a local SH0ES calibration to anchor the background expansion, the parameter space for EDE models becomes unphysically broad, driving the Hubble constant to unfavorably high values. As a result, the EDE-extended models fail to yield a statistically superior fit compared to the standard $\Lambda$CDM cosmology. Ultimately, this demonstrates that the combination of CMB, DESI BAO, and unanchored Pantheon+ datasets robustly prefers a late-time DDE scenario, as recently reported by DESI collaborations \cite{DESI:2024mwx,DESI:2025zgx}.
Finally, in our most comprehensive analysis with the full dataset, the internal tension within the baseline models is maximized. The standard CPL model attempts to accommodate the high-$H_0$ local SH0ES data alongside the CMB observations by driving its EoS parameters further away from the $\Lambda$CDM baseline. On the other hand, both EDE-extended models  accommodate the high local Hubble constant measurement, yielding significantly better global fits. In this complete dataset setting, the EDE-CPL model provides the best overall description, successfully and simultaneously providing the early-universe adjustments needed for the Hubble constant crisis and the late-universe dynamics favored by DESI. In conclusion, we observe that the DDE signal indicated by the DESI BAO data is a late-time phenomenon that cannot be described merely by early-time modifications to the expansion history through EDE.
\end{itemize}

%=================================================================
\section{DATA AVAILABILITY}
The data used in this work are available publicly.
\FloatBarrier
\bibliographystyle{apsrev4-1}
\bibliography{ref}
\end{document}